# Effects of Plasmoid Formation on Sawtooth Process in a Tokamak

A. Ali[1,2] and P. Zhu[1,3,4,a]

[1]*CAS Key Laboratory of Geospace Environment and Department of Engineering and Applied Physics, University of Science and Technology of China, Hefei 230026, China*

[2] *National Tokamak Fusion Program, Islamabad 3329, Pakistan*

[3]*KTX Laboratory and Department of Engineering and Applied Physics, University of Science and Technology of China, Hefei 230026, China*

[4]*Department of Engineering Physics, University of Wisconsin-Madison, Madison, Wisconsin 53706, USA*

For realistic values of Lundquist number in tokamak plasmas, the 1/1 magnetic island leads to the formation of secondary thin current sheet, which breaks up into a chain of small magnetic islands, called plasmoids. The role of plasmoid dynamics during the sawtooth reconnection process in fusion plasmas remains an unresolved issue. In this study, systematic simulations are performed to investigate the resistive internal kink mode using the full resistive MHD equations implemented in the NIMROD code in a simplified tokamak geometry. For Lundquist number $S \geq 1.6 \times 10^7$, secondary current sheet is found to be unstable to plasmoids during the nonlinear resistive kink mode evolution with a critical aspect ratio of the current sheet of ~70. The merging of small plasmoids leads to the formation of a monster plasmoid that can significantly affect the primary island evolution. This may provide an explanation for the partial reconnection observed in sawtooth experiments.

## I. Introduction

Sawtooth crash is a typical example of fast magnetic reconnection observed in tokamak plasmas when the value of safety factor $q$ becomes less than one on the magnetic axis. It is widely believed that the sawtooth crash is triggered by fast growing magnetic island due to the m/n = 1/1 internal kink mode[1,2] (where m and n are the poloidal and toroidal mode numbers, respectively). Experimental results confirm that the internal kink mode play an important role in the onset of sawtooth oscillations[3-5]. The sawtooth process has received continued research interests[6-11] because of its deleterious effects on the tokamak plasma confinement, even though comprehensive understanding of the underlying physics remains missing.

The most simplistic explanation of the sawteeth phenomena was proposed by Kadomtsev[2]. However, the collapse time predicted by Kadomtsev is based on the Sweet-Parker (SP) model[12,13], which is much longer than that observed in the tokamak experiments[14,15]. Therefore, an interpretation of the sawteeth phenomena beyond the Kadomtsev model is required to resolve the discrepancy between theory and observations. It has long been suggested that the SP-like current sheet during the nonlinear evolution of resistive kink mode may become tearing unstable and break into plasmoids. Such plasmoids may provide a possible mechanism for the fast reconnection process during the sawteeth oscillation.

It has been reported by many authors that when aspect ratio of the SP like current sheet exceeds a critical value of ~100, the current sheet can break up and plasmoids can form along the current sheet, leading to faster reconnection[16-25]. In recent simulation studies, Yu *et al.*[26] and Gunter *et al.*[27] find that plasmoid formation can accelerate reconnection during the sawtooth process. Plasmoid formation has also been observed in resistive magnetohydrodynamics (MHD) simulations[28] of the NSTX helicity injection experiment. However, most of these simulations on the plasmoid formation during the sawtooth process have been performed either in two-dimensional (2D) slab configuration or reduced MHD models. Further efforts are needed to explore the plasmoid dynamics in more realistic tokamak configuration and plasma parameter regime with sufficiently high resolution using the resistive MHD model with full set of equations.

In this study, we perform simulation study of the resistive kink mode in a simplified tokamak geometry (i.e. cylindrical geometry) using the full resistive MHD equations implemented in the NIMROD code[29]. To compare our simulation results with the reduced MHD simulations, we select the ASDEX-Upgrade tokamak parameters. Complete evolution of the resistive kink mode from linear to the explosive nonlinear growth phase is analyzed for the onset criteria of plasmoid formation. Both the onset and the dynamics of plasmoids (in particular the central monster plasmoid) are much different from those in previous reduced MHD simulations[26]. Although initially the generation of plasmoids speeds up the reconnection process, these plasmoids eventually merge together into a bigger plasmoid and change the subsequent reconnection process. The monster plasmoid significantly influences the saturation width of the primary magnetic island and thus causes the partial magnetic reconnection, which may provide a possible mechanism for the partial sawteeth crashes observed in tokamak experiments. The new aspects of the plasmoid formation and its effects on the sawtooth process obtained from this study may help the stability analysis of future tokamak experiments such as ITER and CFETR[30-32].

The remaining paper is organized as follows: the numerical model is described in Sec. II. The nonlinear evolution of resistive kink mode for a typical case is illustrated in Sec. III, where the plasmoid formation is demonstrated using Poincare plots and toroidal current density contours. In Sec. IV, we perform a comparison of the kinetic energy evolution, saturation island widths of primary magnetic island and the monster plasmoid for different resistivity cases. Finally in Sec. V, we present a summary and discussion on our study.

## II. MHD Model and Equilibrium

Our results are based on numerical simulations of the single-fluid full MHD model implemented in the NIMROD code[29]. The single-fluid equations can be written as follows:

$$\frac{\partial \rho}{\partial t} + \nabla \cdot (\rho \mathbf{v}) = 0 \tag{1}$$

$$\rho \left( \frac{\partial}{\partial t} + \mathbf{v} \cdot \nabla \right) \mathbf{v} = \mathbf{J} \times \mathbf{B} - \nabla p + \rho \nu \nabla^2 \mathbf{v} \tag{2}$$

$$\frac{N}{(\gamma - 1)} \left( \frac{\partial}{\partial t} + \mathbf{v} \cdot \nabla \right) \mathbf{T} = -p \nabla \cdot \mathbf{v} - \nabla \cdot \mathbf{q} \tag{3}$$

$$\frac{\partial B}{\partial t} = \nabla \times (\mathbf{v} \times \mathbf{B} - \eta \mathbf{J}) \tag{4}$$

$$\nabla \times \mathbf{B} = -\mu_0 \mathbf{J} \tag{5}$$

$$\mathbf{q} = -N(\chi_\parallel \nabla_\parallel \mathbf{T} + \chi_\perp \nabla_\perp \mathbf{T}) \tag{6}$$

where $\rho$, $N$, $p$, $\mathbf{J}$, $\mathbf{B}$, $\mathbf{v}$, $\mathbf{q}$, $\gamma$, $\eta$, $\nu$, $\chi_\parallel$ and $\chi_\perp$ are the plasma mass density, number density, pressure, current density, magnetic field, velocity, heat flux, specific heat ratio, resistivity, viscosity, parallel and perpendicular thermal conductivity, respectively. The Lundquist number is defined as $S = \tau_R / \tau_A$, where $\tau_R = \mu_0 a^2 / \eta$ is the resistive time and $\tau_A = a/v_A$ the Alfvénic time ($a$ is the minor radius). The Alfvén speed $v_A$ is defined as $v_A = B_0/\sqrt{\mu_0 \rho}$. In these simulations, $S$ varies from $10^6$ to $10^8$ with magnetic Prandtl number $Pr = \nu/\eta = 0.1$, which is typical for tokamak plasma. The equilibrium safety factor $q$ profile is

$$q(r/a) = q_0 \left[ 1 + \left( \frac{(r/a)}{\rho_0} \right)^{2\Lambda} \right]^{\frac{1}{\Lambda}} \tag{7}$$

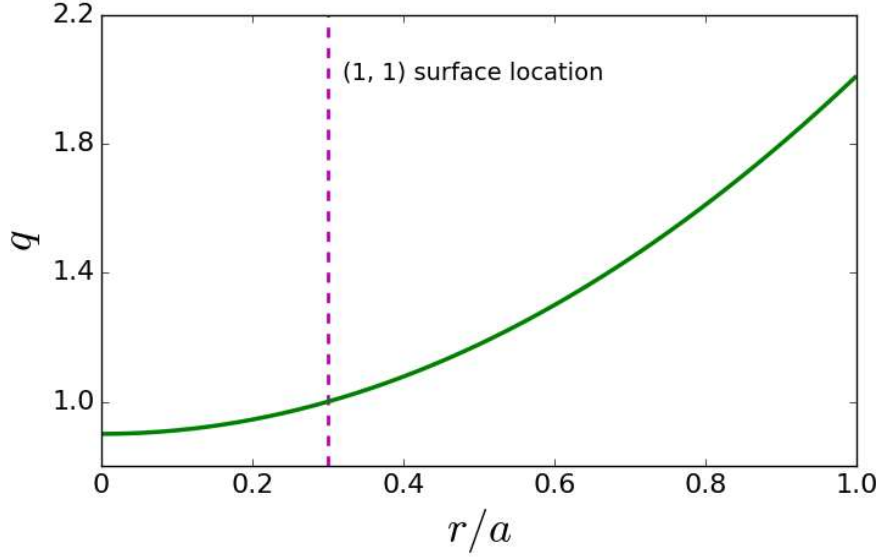

Fig. 1: Radial profile of the equilibrium safety factor.

where $q_0 = 0.9$, $\rho_0 = 0.6$, and $\Lambda = 1$ (Fig. 1). The rational surface (1, 1) is located at the radial position r = 0.3a , as marked by the dashed line.

### III. Simulation Results

Nonlinear simulations are performed to investigate the evolution of resistive kink mode and the associated plasmoids formation based on the single-fluid full MHD equations implemented in the NIMROD code. The simulations are performed in a broad range of plasma resistivity corresponding to $S = 10^6$ to $S = 10^8$ for the same equilibrium introduced earlier. For these simulations, we adopt the ASDEX-Upgrade parameters with aspect ratio $A = 3.3$, minor radius $a = 0.5m$, toroidal magnetic field $B = 2.0$, and number density $n = 3 \times 10^{19} m^{-3}$. The simulation results for a typical case of $S = 2.3 \times 10^8$ are presented. Time evolution of the perturbation kinetic energy and the corresponding growth rate for the total and $n = 1$ toroidal modes are plotted in Fig. 2. The corresponding time development of Poincaré plot, toroidal current density and radial flow contours are shown in Figs. 3-5.

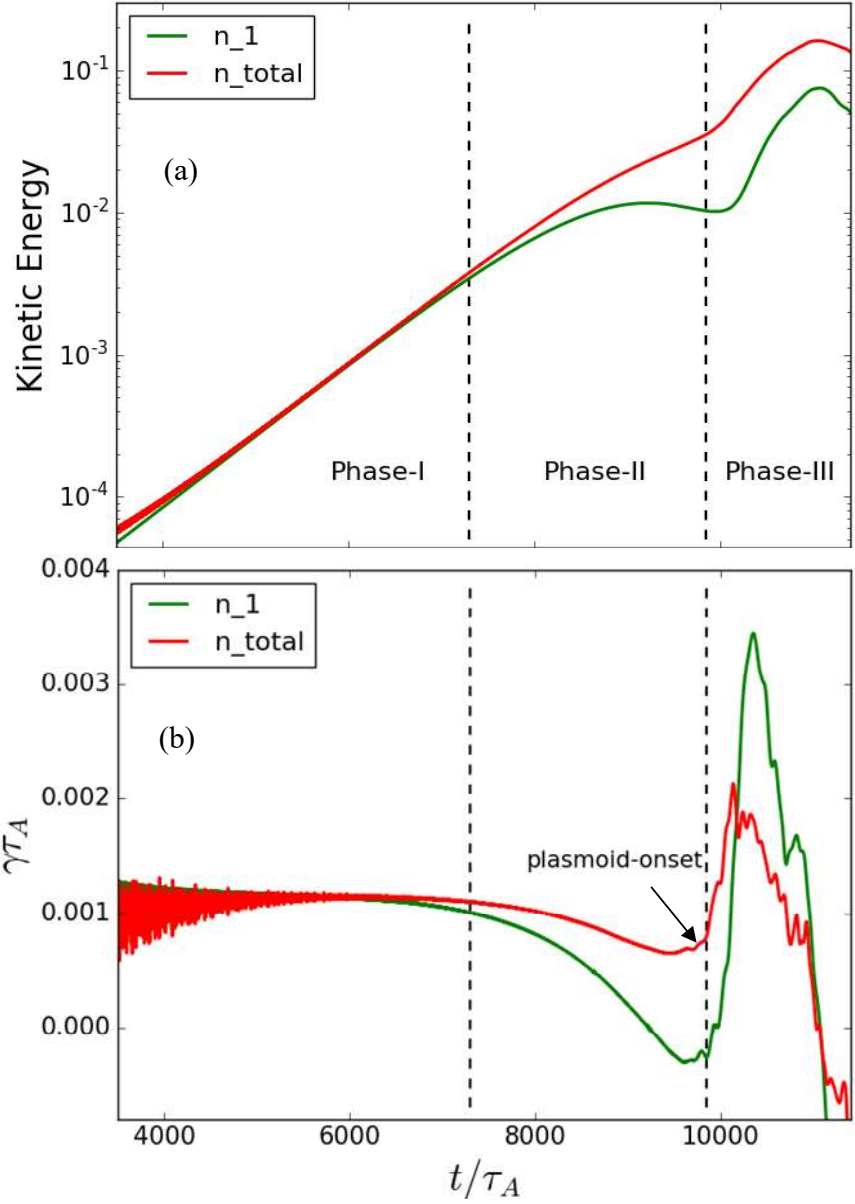

Fig. 2: (a) Time evolution of kinetic energy for the toroidal mode number $n = 1$ and the total perturbation. (b) The instantaneous growth rate of the kinetic energy, which is defined as $\gamma = d[\ln(E)]/dt$.

The complete evolution of the resistive kink mode kinetic energy can be divided into three phases, where the kinetic energy grows linearly in phase-I, forming a thin magnetic island at the rational surface. In phase-II the kinetic energy growth slows down and SP-like current sheet starts to form. The SP-like current sheet continues thinning and elongating along the radial and z-direction, respectively. Such an evolution of the kinetic energy of the resistive kink mode is quite similar to that of the tearing mode[23]. However, it may be noted that the associated magnetic island and secondary SP-like current sheet evolution are significantly different in the two cases. In the case of resistive tearing mode the secondary current sheet is formed as a result of X-point collapse in the nonlinear evolution[23], while in the case of resistive kink mode, the secondary SP-like current sheet is formed in beginning of the nonlinear phase-II.

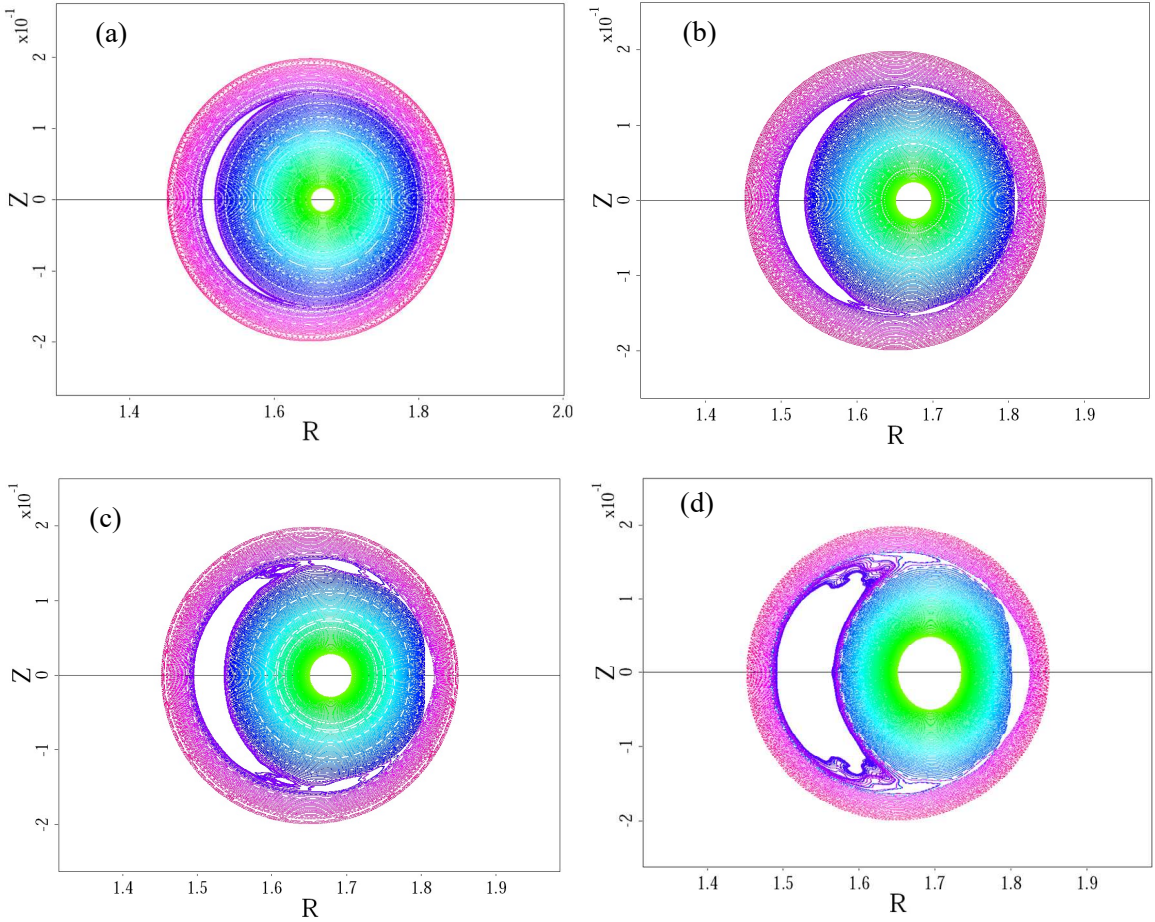

Fig. 3: Poincaré plots of the magnetic field lines at different times: (a) during the SP-like reconnection (Phase-II); (b) during the initial plasmoid unstable stage (Phase-III), where 5 small plasmoids form along the current sheet; (c) when the smaller plasmoids coalesce to form bigger central plasmoid; (d) when the monster plasmoid forms during the saturation stage.

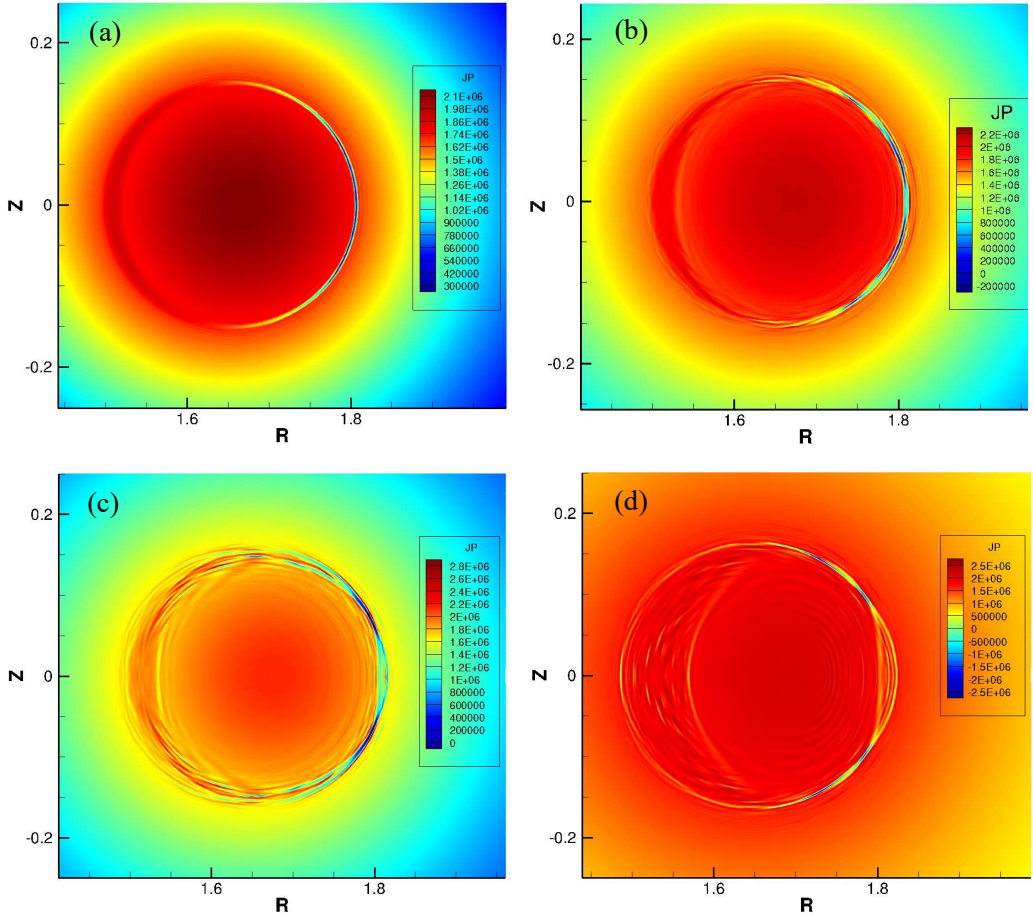

Fig. 4: Toroidal current density contours at different times corresponding to those in Fig. 3: (a) when SP-like secondary current sheet forms; (b) the initial unstable stage of the secondary current sheet, where 5 small plasmoids form and 4 tertiary current sheets emerge; (c) when smaller plasmoids coalesce to form bigger central plasmoid; (d) and when monster plasmoid forms at the final saturation time.

The thinning of SP-like current sheet beyond a critical current sheet aspect ratio results in a faster reconnection process as indicated by the abrupt increase in the kinetic energy growth in Phase-III (Fig. 2(a)). Although the total kinetic energy of the kink mode increases explosively in phase-III (Fig. 2), the amplitude of $n = 1$ is much lower than the total, which signifies the contribution of higher $n$ modes in this phase. The explosive increase in the growth rate is evidently observed after the slow nonlinear growth phase (Fig. 2b), which is due to the formation of plasmoids along the narrow secondary current sheet (Fig. 3).

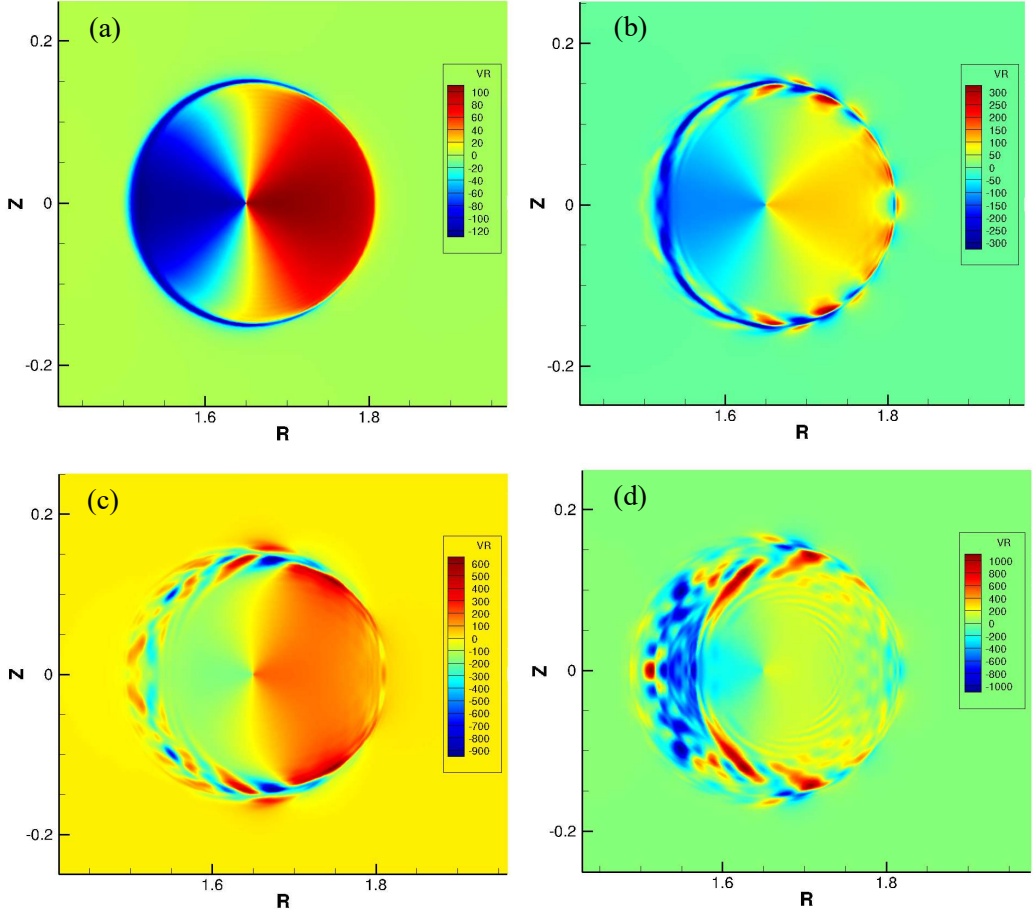

Fig. 5: Contours of the radial plasma flows during the (a) SP-like reconnection phase; (b) initial unstable stage of the secondary current sheet; (c) plasmoid coalescence stage; (d) saturation time.

As the current sheet narrows down (Fig. 3(a)), it becomes susceptible to the plasmoid instability. Consequently, small plasmoids form along the secondary current sheet, which leads to the generation of tertiary narrow current sheets among the plasmoids. The splitting of the secondary current sheet into plasmoids with several smaller current sheets speeds up the reconnection process. For this particular case ($S = 2.3 \times 10^8$), five small plasmoids are formed along the current sheet as depicted in Fig. 3(b). These tiny plasmoids grow with time and merge with each other as they move along the current sheet. These plasmoids eventually form a monster plasmoid at the center of the original current sheet as shown in Fig. 3(c). Such a monster plasmoid can interact with the primary magnetic island and can possibly modify its dynamics. The final quasi-saturation stage with a much bigger central plasmoid is shown in Fig. 3(d). The plasmoids dynamics is further demonstrated in Fig. 4, where the generation of secondary current sheet and its breaking into several small tertiary current sheets are explicated. Another important feature associated with the plasmoid instability is the plasma flow evolution, since the formation of plasmoids may considerably modify the flow evolution inside the $q = 1$ surface (Fig. 5). The contribution of higher modes during the plasmoid-unstable stage is suggested by the localized radial flow patterns (Fig. 5(b) and Fig. 5(c)), which eventually becomes turbulent (Fig. 5(d)).

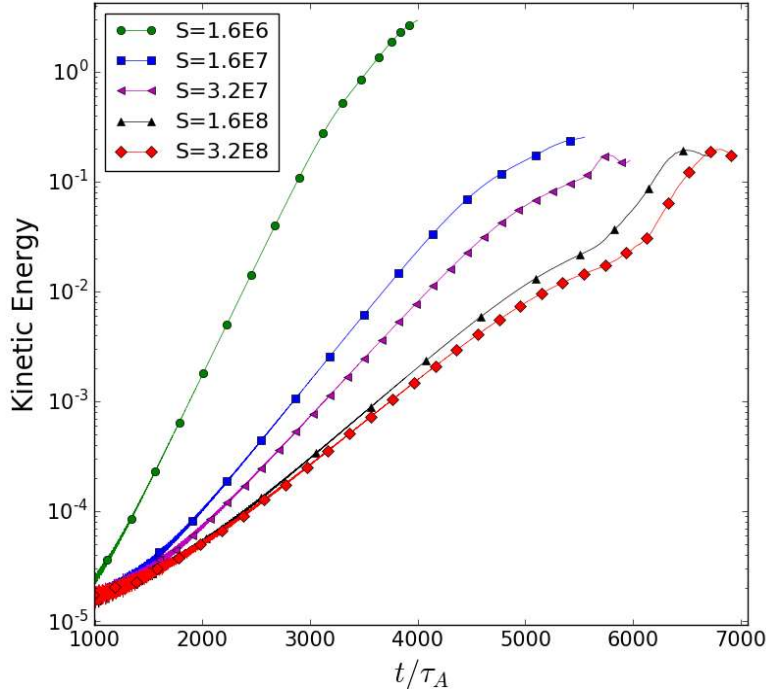

Fig. 6: Time evolution of kinetic energy for five different Lundquist number cases.

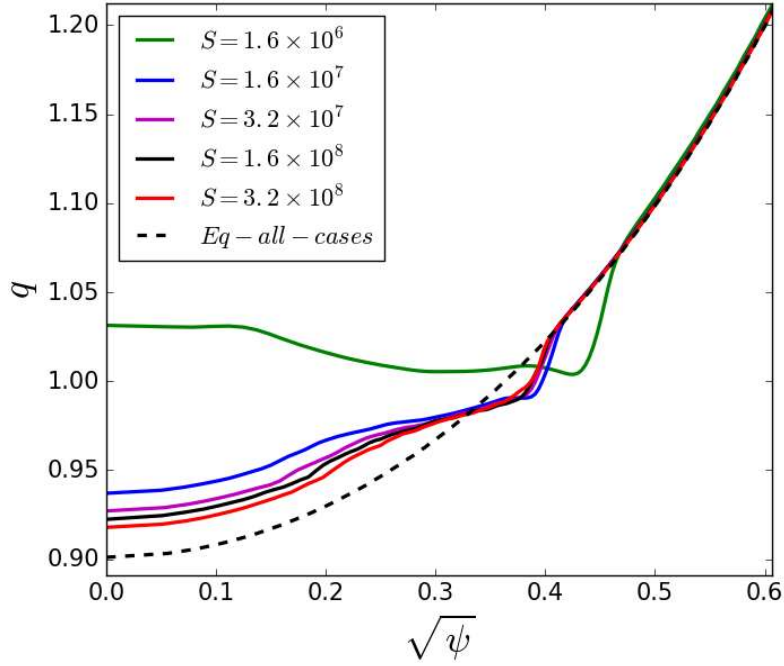

Fig. 7: The safety factor profiles during the saturation phase for five different resistivity cases, corresponding to those in Fig. 6.

IV. **The Impact of Plasmoids on the Sawtooth Reconnection Process**

We further investigate the plasmoid formation and their dynamics in the regimes of $S = 10^6$ to $S = 10^8$ for five different resistivity cases (Fig. 6). The linear growth rate of the resistive kink mode decreases with the increasing value of Lundquist number, hence the plasmoid onset is delayed in the higher $S$ (or lower resistivity) cases. In higher resistivity case ($S = 1.6 \times 10^6$), neither plasmoid formation takes place nor the fast reconnection phase is observed. The kinetic energy saturates early and full reconnection occurs.

The safety factor profile evolution due to nonlinear resistive kink mode can be used to tell whether the reconnection is full or partial (Fig. 7). The original $q$-profile (depicted by the dashed black line in Fig. 7) evolves in two ways; the value of the safety factor on the magnetic axis $q_0$ increases as the reconnection progresses, and the $q$-profile around the (1,1) surface flattens in time. For the higher resistivity case ($S = 1.6 \times 10^6$), the value of safety factor at the magnetic axis ($q_0$) raises above unity, which indicates full reconnection. For lower values of the plasma resistivity ($S = 1.6 \times 10^7$), $q_0$ remains below unity, which suggests that partial reconnection occurs and the final stage is a quasi-saturation state. For this particular case, the plasmoid unstable regime is very brief and a relatively small plasmoid is generated at the center of the secondary current sheet. For $S > 1.6 \times 10^7$, plasmoid instability could be responsible for the explosive growth of kinetic energy shown in Fig. 2, as well as the emergence of a large central plasmoid at the center of the secondary current sheet during the final phase of nonlinear evolution. Such a monster plasmoid interacts with the primary magnetic island and eventually leads to a quasi-saturation state responsible for the partial reconnection.

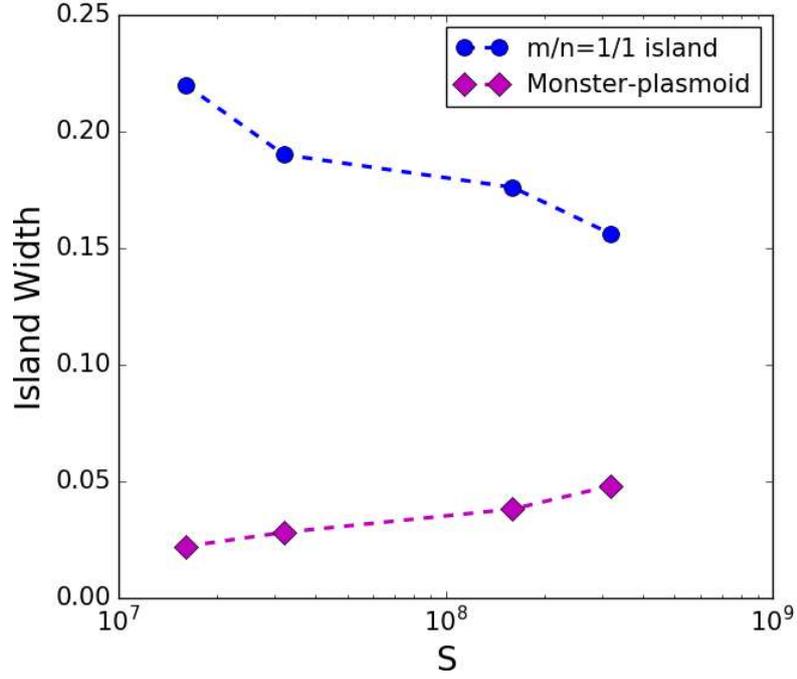

Fig. 8: The saturation width of the primary magnetic island and the monster plasmoid as a function of the Lundquist number, where the island widths are normalized by the plasma minor radius.

The role of plasmoid formation may be further quantified using the primary magnetic island and the monster plasmoid widths at the final saturation time in different resistivity cases (Fig. 8). It is observed that the saturation island width decreases with the Lundquist number, whereas the monster plasmoid width increases. Such a finding demonstrates the impact of the monster plasmoid on the final saturation level of resistive kink mode and may explain the partial reconnection in tokamak sawtooth experiment.

**V. Conclusion and Discussion**

We have performed numerical simulations of the nonlinear resistive kink mode using a single-fluid full resistive MHD model implemented in the NIMROD code for a cylindrical tokamak geometry. The main conclusions of this study are:

  i. The nonlinear reconnection process can be divided into two phases, where the plasmoid dominated explosive reconnection is triggered after the slow SP-like reconnection.

  ii. Small plasmoids form along the current sheet above a critical current sheet aspect ratio of ~70 (corresponding to the case with $S = 1.6 \times 10^7$).

  iii. The small plasmoids coalesce with each other, forming a monster plasmoid.

  iv. The formation of monster plasmoid slows down the reconnection process in the final stage and leads to the partial reconnection.

Our results indicate that the fast reconnection phase is dominated by the plasmoids formation process along the secondary current sheet and followed by a monster plasmoid induced saturation stage. Our results are consistent with the reduced MHD simulations[26] on plasmoid-dominant fast reconnection, however, the plasmoid onset and dynamics are measurably different in our full MHD simulations. In particular, we find opposite dependence of the primary island and the monster plasmoid saturation widths on the plasma resistivity, which is absent in previous results. The key outcome of this study is that the formation of monster plasmoid due to plasmoid instability can significantly affect the sawtooth reconnection process and may provide a mechanism for the partial reconnection that is commonly observed in tokamak sawtooth experiments.

For very high Lundquist numbers $S > 10^9$, the current sheet width decreases to the scale where the two fluid effects become unavoidable and the resistive MHD model is no longer applicable. Therefore, it will be useful to study the plasmoid dynamics using the full MHD model incorporating the two-fluid effects in realistic tokamak configuration. Another important problem that we will consider in future studies is the effect of shear flow on the plasmoid dynamics during the sawtooth process in tokamak plasmas.

# ACKNOWLEDGMENTS

This research was supported by the State Administration of Foreign Experts Affairs-Foreign Talented Youth Introduction Plan under Grant No. WQ2017ZGKX065. We are grateful for the support from the NIMROD team. P. Zhu acknowledges the supports from the National Magnetic Confinement Fusion Science Program of China under Grant No. 2015GB101004, the National Natural Science Foundation of China under Grant Nos. 41474143 and 11775221, and the U. S. Department of Energy under Grant Nos. DF-FG02-86ER53218 and DE-FC02-08ER54975. The computational work used the XSEDE resources (U.S. NSF Grant No. ACI-1053575) provided by TACC under Grant No. TG-ATM070010. This research also used the computing resources from the Supercomputing Center of University of Science and Technology of China.